\documentclass[aps,pra,twocolumn,10pt]{revtex4}

\usepackage{amsmath}
\usepackage{amssymb}
\usepackage{amsthm}
\usepackage[pdftex]{graphicx}
\usepackage{color}

\newcommand{\bra}[1]{\langle {#1} \vert}
\newcommand{\ket}[1]{\vert {#1} \rangle}

\begin{document}

\title{Macroscopicity of quantum superpositions on a one-parameter unitary path in Hilbert space}

\author{T.J. Volkoff}
\email{volkoff@berkeley.edu}
\author{K.B. Whaley}
\affiliation{Berkeley Quantum Information and Computation Center \\ Dept.~of Chemistry, UC Berkeley}


\begin{abstract}
We analyze quantum states formed as superpositions of an initial pure product state and its image under local unitary evolution, using two measurement-based measures of superposition size: one based on the optimal quantum binary distinguishability of the branches of the superposition and another based on the ratio of the maximal quantum Fisher information of the superposition to that of its branches, i.e., the relative metrological usefulness of the superposition. A general formula for the effective sizes of these states according to 
the branch distinguishability measure is obtained and applied to superposition states of $N$ quantum harmonic oscillators composed of Gaussian branches. Considering optimal distinguishability of pure states on a time-evolution path leads naturally to a notion of distinguishability time that generalizes the 
well known orthogonalization times of Mandelstam and Tamm and Margolus and Levitin. We further show that the distinguishability time provides a compact operational expression for the superposition size measure based on the relative quantum Fisher information. By restricting the maximization procedure in the definition of this measure to an appropriate algebra of observables, we show that the superposition size of, e.g., N00N states and hierarchical cat states, can scale linearly with the number of elementary particles comprising the superposition state, implying precision scaling inversely with the total number of photons when these states are employed as probes in quantum parameter estimation of a 1-local Hamiltonian in this algebra.
\end{abstract}

\pacs{42.50.Ex, 42.50.-p}

\maketitle

\section{\label{sec:intro}Introduction}
The development of measures of macroscopicity for 
both quantum superposition states \cite{leggett,cirac,bjork,whaleyjan,marquardt,jeong,dur,hornbergersize,hornbergernature} and arbitrary quantum states \cite{dur} has resulted in 
physically motivated orderings of quantum mechanical states based on 
notions of effective size.
Comparisons among measures of state macroscopicity and superposition size \cite{frowislink} reveals that mathematically (and conceptually) closely-related measures give equivalent orderings, while mathematically disparate measures usually require a case-by-case comparison. Measures of macroscopicity are crucial for determining the extent to which quantum states of physical systems exhibiting large quantities of 
either mass, spatial extent, or average number of elementary particles or mode excitations, can be exploited for use as a quantum resource in 
metrology \cite{giovannetti2004quantum}, probing the validity of quantum mechanics in macroscopic settings \cite{hornbergernature} and pure state quantum information processing \cite{jozsa2003role}. Macroscopic superposition states play a prominent role in these situations.  For example, macroscopic quantum superpositions are useful for performing Heisenberg limited metrology \cite{wineland,giovannetti2006quantum,munro2002weak} and for quantum computation \cite{infoprocesscat,mirrahimi2014dynamically}.
If a system exhibits macroscopic quantum behavior in an experiment, an appropriate measure of macroscopicity 
should reveal this behavior. However, because many superposition size measures are defined by an optimization over a subset of observables or set of positive operator-valued measures (POVMs), they can be difficult to calculate, especially for infinite dimensional Hilbert spaces. 

In this paper, we consider quantum superposition states of the following form: \begin{equation}
\ket{\Psi} = {(\mathbb{I}+V)\ket{\Phi}\over \sqrt{2 + 2\text{Re}(z^{N})}}
\label{eqn:state}\end{equation} where $\ket{\Phi} := \ket{\phi}^{\otimes N}$ is a product state in the $N$-th tensor product of a single-mode Hilbert space $\mathcal{H}$, $V:=U^{\otimes N}$ is a tensor product of the same local unitary operator, and $z:= \langle \phi \vert U \vert \phi \rangle $. 
The unitary operator $V$ can be considered as the exponential of a 1-local, self-adjoint operator $\theta \sum_{i=1}^{N}h^{(i)}$ with $h^{(j)}=h=h^{\dagger}$ acting only on mode $j$ for all $j$, i.e., a 1-local observable, and $\theta$ a real number parametrizing the evolution. We assume the dimension of the single-particle Hilbert space is at most countably infinite. For $N$ finite, the state $\ket{\Psi}$ can be considered as a generalization of the well-known GHZ-like states with nonorthogonal branches that are defined on the finite dimensional space $(\mathbb{C}^{2})^{\otimes N}$ \cite{cirac} to 
the corresponding states in countably infinite dimensional Hilbert spaces.  

The class of states with form $\ket{\Psi}$ contains many equal-amplitude superposition states that have been previously characterized as macroscopic. For example, 
 the $N$-mode GHZ (Greenberger-Horne-Zeilinger) state defined in $\mathbb{C}^{2})^{\otimes N}$ \cite{greenberger1989going} can be written in the general form $\ket{\Psi}$ by using $\ket{\phi} = \ket{0}$ and $U=\sigma_{x}$. Certain entangled bosonic states (defined in the appropriate symmetric subspace of $\ell^{2}(\mathbb{C})^{\otimes N}$)  can also be written in this way. For example, the N00N state $\propto \ket{0}\ket{M} + \ket{M}\ket{0}$ ($M\in \mathbb{Z}_{\ge 0}$) can be obtained by applying the unitary operator $\mathbb{I}\otimes (\ket{0}\bra{M} + \ket{M}\bra{0})$ to the state $\ket{\Psi}$ with $N=2$, $\ket{\phi} =\ket{0} $ and $U= \ket{0}\bra{M} + \ket{M}\bra{0}$. 
The entangled even coherent states of the electromagnetic field, which are proportional to $\ket{\alpha}^{\otimes N} + \ket{-\alpha}^{\otimes N}$ \cite{volkoff}, can be written by using $\ket{\phi} = \ket{\alpha}$ and $U=\text{exp}(-i\pi a^{\dagger}a)$. 
Superpositions of the form $\ket{\Psi}$ also appear in quantum condensed matter contexts. For example, interference between two weakly-interacting Bose-Einstein condensates can be described by considering the order parameter field as a single mode and evaluating the sinusoidal term in the condensate density for a superposition of two order parameter fields with nearly disjoint supports \cite{pitaevskii}. In the theory of two-dimensional quantum critical phenomena, one can consider the basis states as eigenstates $\ket{ [ \varphi(x)]}$ of the quantum field operator so that the single-mode Hilbert space is now a space of sufficiently smooth complex-valued functions (the field operator is obtained from, e.g., canonical quantization of the Euler-Lagrange equation at the Lifshitz point). With this picture, it can be shown that in the quantum Lifshitz model (describing, e.g., 2-D quantum dimers \cite{ardonnefradkin}) a normalized superposition of the ground state and a single vortex having circulation $m \in \mathbb{Z}$ is also of the form $\ket{\Psi}$.

In this work, we 
focus primarily on states $\ket{\Psi}$ in which $\ket{\phi}$ and $U\ket{\phi}$ are non-orthogonal, although we shall also briefly discuss interesting examples of superpositions with orthogonal branches. 
Analysis of GHZ-like states usually treats $z$ merely as a state parameter determining the overlap between single mode states $\ket{\phi_1}$ and $\ket{\phi_2} = U \ket{\phi_{1}}$, without consideration of the role of the unitary transformation $U$ in relating these states. 
Here we shall determine the effective size of $\ket{\Psi}$ for a general $U$ with two measures of size, i) as measured by branch distinguishability \cite{whaleyjan} (labeled $C_{\delta}$) and ii) as measured by metrological usefulness of the superposition \cite{dur} (labeled $N^{rF}$), and analyze the implications of this analysis for two classes of unitary operators.  In the first case we consider unitaries that cause quadrature squeezing and/or displacement of optical fields. Our analysis for this class of states shows that for a state $\ket{\Psi}$ defined with a value of the inner product $z$ that decays exponentially with some
physical quantity
(e.g., for photonic states, a squeezing parameter or photon number), the measures of superposition size $C_{\delta}$ and $N^{rF}$
both exhibit a linear dependence  on 
that same quantity.  

The second case we consider is the time evolution operator $U = e^{-iHt/\hbar}$.  Here the analysis of branch distinguishability for Eq. (\ref{eqn:state}) requires a consideration of the time evolution of macroscopically distinct states that leads to a generalization of the Mandelstam-Tamm \cite{mandelstamtamm} and Margolus-Levitin \cite{margolus} energy-time inequalities to non-orthogonal states.

In this paper we focus exclusively on the $C_{\delta}$ and $N^{rF}$  superposition size measures because for discrete (spin) systems, each of these serves as a representative of a 
class of measures whose elements assign the same scaling of superposition size for a given superposition.
The class represented by the metrological usefulness measure $N^{rF}$ is contained in the class represented by $C_{\delta}$ \cite{dur} because macroscopicity according to $N^{rF}$ implies macroscopicity according to $C_{\delta}$ under physically reasonable constraints  \cite{frowislink}.
In this work, we show that the general state in Eq.(\ref{eqn:state}) is macroscopic according to $C_{\delta}$ if and only if it is macroscopic according to $N^{rF}$. 

The remainder of the paper is structured as follows.  In Section \ref{sec:size}, we derive an expression for the branch distinguishability superposition size for states of the form $\ket{\Psi}$ in an arbitrary  
countably infinite tensor product Hilbert space and demonstrate its usefulness by comparison of the sizes of various superpositions of Gaussian product states.  Section \ref{sec:time} treats the important case of superpositions of states lying along a given unitary time-evolution path in state space. Here, in addition to relating the rate of change of branch distinguishability superposition size to the Fubini-Study line element on the path defining $\ket{\Psi}$, we 
generalize the well known orthogonalization times of Mandelstam-Tamm \cite{mandelstamtamm} and Margolus-Levitin to answer the question ``What is the minimal time that must elapse for a given pure quantum state to evolve to a state from which it is optimally distinguishable with a predetermined success probability?'' This leads to the definition of a minimal distinguishability time, which is then used in Section \ref{sec:variance} to show that the metrologically-motivated measure $N^{rF}$ of superposition size \cite{dur} may be expressed operationally in terms of the minimal distinguishability times of the superposition $\ket{\Psi}$ and its branches.  For spin systems in $(\mathbb{C}^{2})^{\otimes N}$, we can explicitly construct a local observable which guarantees a large $N^{rF}$ value for $\ket{\Psi}$ as long as its branches are nearly orthogonal. This construction shows that the superposition sizes measured by $C_{\delta}$ and $N^{rF}$ are equivalent for states $\ket{\Psi}$ having $\vert z \vert \ll 1$. This analysis is then generalized in Section \ref{sec:photons} to the case of oscillator systems in 
$\ell^{2}(\mathbb{C})^{\otimes N}$, by constructing an algebra of observables containing an element 
that exhibits an extensive difference between the variance in the superposition state and in its branches.  We show that this ``metrological macroscopicity algebra'' allows several examples of 
multimode photonic superpositions to exhibit a value of $N^{rF}$ scaling linearly with the 
average number of photons comprising the state. As a consequence, it is in principle possible to obtain
extensive improvements in the maximal precision of parameter estimation beyond the traditional Heisenberg bound of $\sim {1\over N}$ in a Hilbert space $\mathcal{H}^{\otimes N}$. Finally, in Section \ref{sec:conc} we summarize and conclude.

\section{\label{sec:size}General branch distinguishability superposition size}
For a superposition state of the form $\ket{\psi}=\ket{A} + \ket{B} \in  \mathcal{H}^{\otimes N}$, the 
branch-distinguishability superposition size $C_{\delta}(\ket{\psi})$ is defined by \cite{whaleyjan} \begin{equation} C_{\delta}(\ket{\psi}) = {N\over n_{\text{eff}}(\delta , \ket{\psi})}\label{eqn:disting}
\end{equation} where \begin{equation} n_{\text{eff}}(\delta, \ket{\psi}) = \text{min}\big\lbrace n\big\vert {1\over 2} +{1\over 4}\Vert \rho_{A}^{(n)} - \rho_{B}^{(n)} \Vert_{1} \ge 1-\delta \big\rbrace \label{eqn:helstrom} \end{equation} in which $\Vert \cdot \Vert_{1}$ signifies the trace norm on bounded operators on $\mathcal{H}^{\otimes n}$ and $\rho_{A}^{(n)} :=\text{tr}_{N- n}\ket{A}\bra{A}$ (\textit{mutatis mutandis} for $\ket{B}$) is the $n$-reduced density matrix ($n$-RDM) of $\ket{A}\bra{A}$. 

$C_{\delta}(\ket{\psi})$ constitutes a measurement-based measure of superposition size that is based on the notion that branches of the more macroscopic superpositions can be distinguished with maximal probability by measurements of subsystems of smaller size.
To see this, note that the expression on the left hand side of the inequality in Eq.(\ref{eqn:helstrom}) is the maximal probability $p_{\text{H},\text{succ}}(\rho_{A}^{(n)},\rho_{B}^{(n)})$ over all possible $n$-mode measurements of successfully distinguishing the $n$-RDM corresponding to the branches $\ket{A}\bra{A}$ and $\ket{B}\bra{B}$ \cite{helstromone}. This fact leads us to the following interpretation: given $\delta \in (0,1/2)$, $C_{\delta}(\ket{\psi})$ is the largest number $N\over n_{\text{eff}}(\delta)$ of subsystems existing in a state of superposition such that an optimal $n_{\text{eff}}(\delta)$-mode measurement can be used to distinguish the two branches of $\ket{\Psi}$ with probability of success equal to $1-\delta$. It was shown in Ref.~\cite{volkoff} that in order to collapse an equal amplitude, two-branch quantum superposition via measurement, it is sufficient to apply the optimal measurement for distinguishing the two branches of the superposition to the superposition itself.

This measurement-based measure of superposition size has been applied to superposition states of phases of an interacting BEC in a double well potential \cite{whaleyjan}, to electronic supercurrents in a superconducting flux qubit \cite{korsbakken2009}, and to coherent states of a finite chain of electromagnetic cavities, i.e., to photonic superposition states \cite{volkoff}. 

Eq.~(\ref{eqn:disting}) can be applied to the state $\ket{\Psi}$ by first defining an orthonormal basis $\lbrace \ket{e_{1}} = \ket{\Phi} , \ket{e_{2}} ={ V\ket{\Phi} - z^{N}\ket{\Phi} \over \sqrt{1 - \vert z \vert^{2N}}} \rbrace$ of a 2-dimensional subspace of $\mathcal{H}^{\otimes N}$. Transforming the $n$-RDMs $\text{tr}_{N- n}\ket{\Phi}\bra{\Phi}$ and $\text{tr}_{N- n}V\ket{\Phi}\bra{\Phi}V^{\dagger}$ to this basis, we find the following formula for the maximal probability of successfully distinguishing the $n$-RDMs of the branches of $\ket{\Psi}$ by the results of an $n$-mode POVM: \begin{equation}
p_{\text{H},\text{succ}}(\text{tr}_{N-n}\ket{\Phi}\bra{\Phi} , \text{tr}_{N-n}V\ket{\Phi}\bra{\Phi}V^{\dagger} ) = {1\over 2} +{1\over 2}\sqrt{1-\vert z\vert^{2n}}
\label{eqn:rdmprob}\end{equation} with $z$ defined as in Eq.(\ref{eqn:state}). Finally, one obtains $n_{\text{eff}}(\delta)$ by minimizing $n$ to saturate the inequality in Eq.~(\ref{eqn:helstrom}). The result is the general formula: \begin{eqnarray} C_{\delta}(\ket{\Psi}) &=&  {2 \log{\vert \langle \Phi \vert V \vert \Phi \rangle \vert} \over \log(4\delta - 4\delta^{2}) } \nonumber \\ &=& {2N \log{\vert z \vert} \over \log(4\delta - 4\delta^{2}) }
\label{eqn:general}
\end{eqnarray}
It is clear that if $\vert z \vert$ scales exponentially with some quantity that the superposition size $C_{\delta}$ will scale linearly with respect to that quantity. For infinite dimensional Hilbert spaces, this allows for  values of $C_{\delta}$ much larger than $\mathcal{O}(N)$ (even for commonly encountered superpositions) than for spin systems, for which the number of modes $N$ is the natural ``size'' parameter.

There are two important subtleties which must be taken into account when using Eq.~(\ref{eqn:general}). The first is that given a number of modes, $N$, the precision $\delta$ cannot be chosen independently while still maintaining a physically reasonable value of the effective size $n_{\text{eff}}(\delta)$. Given $N \in \mathbb{Z}_{+}$ defining $\ket{\Psi}$, we must have $n_{\text{eff}}(\delta) \in \lbrace 1 , 2, \ldots ,N \rbrace$. For the state $\ket{\Psi}$, this constrains the precision $\delta$ in the following way: \begin{equation}
\delta \in \left( {1\over 2} - {1\over 2}\sqrt{1-\vert z \vert^{2N}}, {1\over 2} - {1\over 2}\sqrt{1-\vert z \vert^{2}} \right) \end{equation} On the other hand, given the precision $\delta \in (0, 1/2)$, one obtains the general formula $n_{\text{eff}}(\delta)=\lceil \log (4\delta - 4\delta^{2})/2\log \vert z \vert \rceil $, so that in order to obtain a small value of $\delta$, the number of modes, $N$, must be very large (at least $n_{\text{eff}}(\delta)$.

The second subtlety is the indeterminate value of $C_{\delta}$ obtained
when $z=0$, i.e., a superposition $\ket{\Psi}$ with $\ket{\phi}$ orthogonal to $U\ket{\phi}$, which, for $(\mathbb{C}^{2})^{\otimes N}$, corresponds to a GHZ state. For these states, the only reasonable choice of the precision is $\delta = 0$, since the reduced density matrices of the branches at any order can be optimally distinguished with unit probability. In physical terms, this means that states of the form $\ket{\psi_{1}}^{\otimes N}$ and $\ket{\psi_{2}}^{\otimes N}$ with $\langle \psi_{1} \vert \psi_{2} \rangle = 0 $ can be equally optimally distinguished by performing a measurement of a single elementary particle, or by performing measurements of any larger subset of particles. For this reason, the GHZ state $ \ket{\text{GHZ}_{N}} \propto \ket{0}^{\otimes N} + \ket{1}^{\otimes N}$ has been defined to have $n_{\text{eff}}=1$, so that 
$C_{\delta = 0} = N$ \cite{korsbakken2009}.  According to the above arguments, there is good reason to generalize this definition to states of the form of $\Psi$ having orthogonal branches in general Hilbert spaces by defining $C_{0}$ to be equal to the total number of \textit{elementary particles} in the state.
(By ``elementary particle'' is simply meant the most fundamental observable particle in the given experiment or energy scale~\cite{whaleyjan}, since $C_{\delta}$ is defined in a measurement-based way.) 
This generalized definition is also necessary for relating the
metrological usefulness measure of superposition size ($N^{rF}$),
 introduced for finite dimensional spin-1/2 systems (having Hilbert space $(\mathbb{C}^{2})^{\otimes N}$) in Ref.\cite{dur} and
extended to countably infinite dimensional systems (having Hilbert space $\ell^{2}(\mathbb{C})^{\otimes N}$) in \cite{volkoff}, to the branch distinguishability measure 
of Eq.(\ref{eqn:disting}). In Section \ref{sec:variance} below, we shall show that if $N^{rF}$ for $\ket{\Psi}$ scales linearly with the total number of elementary particles, then $C_{\delta}$ defined in this generalized sense and given by Eq.~(\ref{eqn:general}) will scale in the same way (and \textit{vice versa}).

The examples of states introduced in Section \ref{sec:intro} provide a testing ground for the branch distinguishability superposition size formula Eq.(\ref{eqn:general}) for $\ket{\Psi}$. For example, consider the following photonic superpositions of $N$-mode Gaussian states
that can be obtained by application of mode squeezing operators and mode displacement operators: \begin{widetext} \begin{eqnarray} \ket{\Psi_{0}} &=& { \left( \bigotimes^{N}_{i=1}S_{i}(-\xi)D_{i}(\alpha) + \bigotimes^{N}_{i=1}S_{i}(\xi)D_{i}(-\alpha) \right) \ket{0}^{\otimes N}  \over A_{0}} \nonumber \\ \ket{\Psi_{1}} &=& { \left( \bigotimes^{N}_{i=1}S_{i}(\xi)D_{i}(\alpha) + \bigotimes^{N}_{i=1}S_{i}(-\xi)D_{i}(-\alpha) \right) \ket{0}^{\otimes N} \over A_{1}}    \nonumber \\  \ket{\Psi_{2,\pm}} &=& {\left( \bigotimes^{N}_{i=1}D_{i}(\alpha)S_{i}(\xi) + \bigotimes^{N}_{i=1}D_{i}(-\alpha)S_{i}(\pm \xi) \right) \ket{0}^{\otimes N} \over A_{2,\pm} } \label{eqn:photonstates}\end{eqnarray}
with $S(\xi) := \exp ( {\overline{\xi}a^{2} - \xi a^{\dagger 2} \over 2})$ the single-mode squeezing operator, $D(\alpha) = \exp (\alpha a^{\dagger} - \overline{\alpha} a )$ the single-mode displacement operator. The branches of these superpositions are products of Gaussian states, i.e., quantum states of an electromagnetic mode having Wigner functions which are Gaussian distributions on the complex plane. In these states, the squeezing parameter $\xi$ and the coherent state parameter $\alpha$ are independent. Taking $\xi, \alpha \in \mathbb{R}_{+}$ for simplicity, the normalization constants are $A_{0} =\sqrt{2+2 \cosh^{-N\over 2}(2\xi)e^{-N{\alpha^{2}\over 2}(1+e^{-2\xi})^{2}(1+\tanh 2\xi)}  } $, $A_{1} =\sqrt{2+2 \cosh^{-N\over 2}(2\xi)e^{-N{\alpha^{2}\over 2}(1+e^{2\xi})^{2}(1-\tanh 2\xi)}  }$ for the two-photon coherent state superposition $\ket{\Psi_{1}}$, and $A_{2,+} = \sqrt{2+2e^{-2N\alpha^{2}e^{2\xi}}}$ and $A_{2,-} = \sqrt{2+2 \cosh^{N\over 2}(2\xi) e^{-2N\alpha^{2}e^{-2\xi}(1-\tanh 2\xi)} }$ for the superpositions of ideal squeezed states $\ket{\Psi_{2,\pm}}$ \cite{mandel}. \end{widetext} 

The superpositions of Eq.(\ref{eqn:photonstates}) are clearly of the form $\ket{\Psi}$ up to a product unitary operator, e.g., \begin{eqnarray}
\ket{\Psi_{2 , +}} &\propto &(\mathbb{I} + (S(-\xi)D(-2\alpha)S(\xi))^{\otimes N})\ket{0}^{\otimes N} \\ \ket{\Psi_{2 , -}} &\propto &(\mathbb{I} + (S(-\xi)D(-2\alpha)S(-\xi))^{\otimes N})\ket{0}^{\otimes N}.
\label{eqn:examplepsi}\end{eqnarray}  From Eq.(\ref{eqn:general}) it is clear that the branch distinguishability superposition sizes of these states can be extracted from a calculation of the normalization constants of the superpositions, because such a calculation requires solving for the vacuum matrix element $z = \langle 0 \vert U \vert 0 \rangle $ in each case. This yields the following superposition sizes: \begin{eqnarray}
\tilde{C}_{\delta}(\ket{\Psi_{0}}) &= & N (\alpha^{2}(1+e^{2\xi})^{2}(1+\tanh 2\xi) \nonumber \\ &{}&- \log \cosh(2\xi) ) \nonumber \\
\tilde{C}_{\delta}(\ket{\Psi_{1}}) &= & N (\alpha^{2}(1+e^{-2\xi})^{2}(1-\tanh 2\xi) \nonumber \\ &{}&- \log \cosh(2\xi) )   \nonumber \\
\tilde{C}_{\delta}(\ket{\Psi_{2,+}}) &= & {N\alpha^{2}e^{2\xi} } \nonumber \\ 
\tilde{C}_{\delta}(\ket{\Psi_{2,-}}) &= & N(4\alpha^{2}e^{-2\xi}(1-\tanh 2\xi)\nonumber \\ &{}&- \log \cosh(2\xi) ) 
\label{eqn:squeezedcatsizes}
\end{eqnarray}
where we have used $\tilde{C}_{\delta} = - C_{\delta}\log(4\delta - 4\delta^{2})$ to put all dependence on the precision $\delta$ on the left hand side. From Eq.(\ref{eqn:squeezedcatsizes}), it is clear that although all the superposition sizes scale linearly with the square modulus of the coherent state amplitude when $\xi = 0$, nonzero quadrature squeezing has different effects on the different states. Specifically, a large 
squeezing parameter $\xi$ results in exponential growth of the superposition size for $\ket{\Psi_{0}}$ and 
$\ket{\Psi_{2,+}}$, while for $\ket{\Psi_{1}}$ and $\ket{\Psi_{2,-}}$, large $\xi$ destroys the dependence of the superposition size on $\vert \alpha \vert^{2}$ and results in linear scaling with the squeezing parameter. 
In contrast to the simple structure of the individual branches, the mathematical expressions for the superposition sizes are surprisingly complicated. We show in the following that for Gaussian branches, a qualitative analysis of the phase space structure of the quasiprobability distributions (which could be given more quantitatively by, e.g., the Husimi $Q$-function) nevertheless allows for a simple analysis of the branch distinguishability superposition size and its dependence on the single mode squeezing and displacement parameters.

\begin{figure}
\includegraphics[scale=.35]{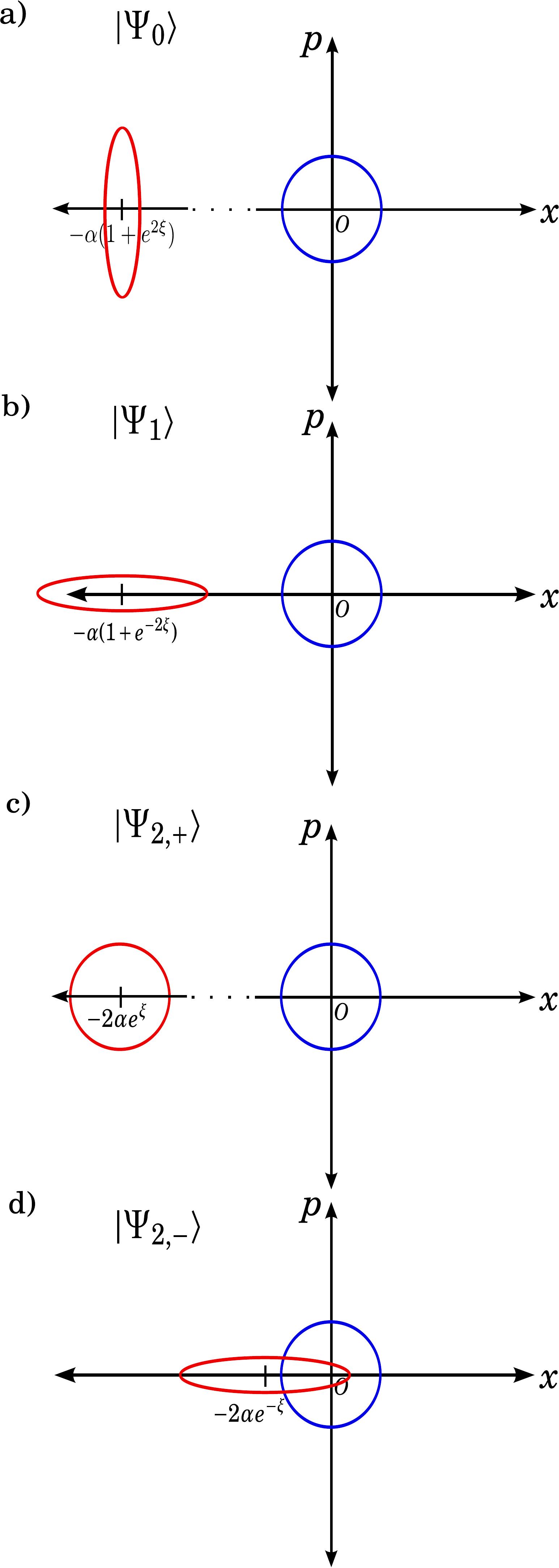}
\vspace{-0.2cm}
\caption{\label{fig:phasespace}Qualitative phase space representations of the variances of $x$ and $p$ quadratures in the vacuum branches (blue) and displaced branches (red) of the states $\ket{\Psi_{0}}$, $\ket{\Psi_{1}}$, $\ket{\Psi_{2,\pm}}$ written in the form $\propto (\mathbb{I} + U)\ket{0}$ for $N=1$ (axes are not to scale). The semimajor and semiminor axes of the ellipses in a), b), and d) are equal to $e^{4\xi}$ and $e^{-4\xi}$ respectively, corresponding to the variances of the respective quadratures. a) $U= S(2\xi)D(-\alpha(1+e^{2\xi}))$ , b) $U=S(-2\xi)D(-\alpha(1+e^{2\xi}))$, c) $U = D(-2\alpha e^{\xi})$, d) $U=S(-2\xi)D(-2\alpha e^{-\xi})$ with $\xi$ taken to be large enough that the squeezed branch quadrature ellipse overlaps that of the vacuum branch.}
\end{figure}

We first note that for real values of $\alpha$ and $\xi$, the adjoint action of the squeezing operator on the displacement operator is \begin{equation} S^{\dagger}(\xi)D(\alpha)S(\xi) = D(\alpha\text{exp}(\xi)) .\end{equation} Applying this relation to Eq.(\ref{eqn:examplepsi}) produces: \begin{eqnarray}
\ket{\Psi_{2 , +}} &\propto &(\mathbb{I} + D(-2\alpha e^{\xi})^{\otimes N})\ket{0}^{\otimes N} \\ \ket{\Psi_{2 , -}} &\propto &(\mathbb{I} + (S(-2\xi)D(-2\alpha e^{-\xi}))^{\otimes N})\ket{0}^{\otimes N}.
\end{eqnarray}  This simplification allows one to easily visualize the branches of these superpositions in phase space and to gain an intuitive, quasi-classical picture of the branch distinguishability superposition size. We illustrate this in Figure \ref{fig:phasespace}, which 
provides a qualitative representation of the quadrature ellipses of the two branches of each of the states in Eq.(\ref{eqn:photonstates}). The blue circles represent the (Fock) vacuum branch, i.e., the coherent state $\ket{0}$, for which all quadratures have equal variance.  Using dimensionless quadrature amplitudes $\hat{x}, \hat{p}$ with commutation relation $[\hat{q},\hat{p}] = i$, the radius of this branch is equal to ${1\over 2}$.
The red ellipses represent the variances of $\hat{x}$ and $\hat{p}$ for the displaced and/or squeezed branches.  Thus, for the displaced (red) branch of $\ket{\Psi_0}$  the semi major axis is given by the $\hat{p}$ quadrature variance $e^{4\xi}$ and the semi minor axis by the $\hat{x}$ quadrature variance $e^{-4\xi}$. These ellipses could also be visualized as appropriately chosen level curves of phase space quasiprobability densities (e.g., $P$-, $W$-, or $Q$-functions \cite{harochebook}) for the branches of the superpositions.  Eq.(\ref{eqn:squeezedcatsizes}) shows that the superposition sizes of these states exhibit different behaviors as the magnitude of the squeezing parameter $\xi$ goes to infinity. The plots in Fig. \ref{fig:phasespace} indicate that one may associate the appearance of asymptotically exponential scaling of $C_{\delta}$ in the squeezing parameter $\xi$ with the absence of overlap between the squeezed branch ellipse and the coherent state circle as $\xi \rightarrow \infty$. Thus, in panels a) and c), the quadrature ellipses do not asymptotically overlap as $\xi \rightarrow \infty$, while in panels b) and d) they do (note that for panel d) we have assumed that $\xi$ is large in order to illustrate that the centers of the quadrature ellipses coincide as $\xi \rightarrow \infty$).

If the squeezing parameter $\xi$ is taken to be dependent on the displacement parameter $\alpha$, more interesting functional forms are encountered for the scaling of superposition size of the states in Eq.(\ref{eqn:photonstates}). In particular, it is evident from the superposition size of $\ket{\Psi_{2,+}}$ in Eq.(\ref{eqn:squeezedcatsizes}) that for the special case $\xi =k \log \alpha \in \mathbb{R}$, where $\xi$ scales logarithmically with $\alpha$, a polynomial scaling $\mathcal{O}(\alpha^{k+1})$ of the branch distinguishability superposition size with the displacement $\alpha$ can be obtained.  Finally, we note that it is also possible to generate exponential superposition size scaling using only a logarithmic increase of the squeezing parameter, independent of the displacement. For example, the superposition $\ket{\Psi}$ with $U = \prod_{k=0}^{\infty}S(-\log x)^{k}D({\alpha \over k!})S(\log x)^{k}$ and $\ket{\phi} = \ket{0}$ gives a value of $C_{\delta}$ in $\mathcal{O}(\alpha e^{x})$, where $\xi = \log x$ and is independent of $\alpha$. From this example, one sees that superexponential scaling $\mathcal{O}(\alpha e^{e^{\xi}})$ of the superposition size of $\ket{\Psi}$ can be obtained by taking $U=\prod_{k=0}^{\infty}S(-\xi)^{k}D({\alpha \over k!})S(\xi)^{k}$ and $\ket{\phi} = \ket{0}$.

\section{\label{sec:time}Case of $U = e^{-iHt\over \hbar}$}

We now consider the situation when the unitary $V$ appearing in $\ket{\Psi}$ of Eq.(\ref{eqn:state}) is the unitary time evolution operator,
i.e., $V = \text{exp}(-{it\over \hbar}\sum_{i=1}^{N}h^{(i)})$ where $h^{(j)}=H$ for all $j$. This allows analysis of superpositions of states lying on a path $\mathcal{C} := \lbrace U(t)\ket{\phi} \vert 0\le t \le T \rbrace$ where $U(t)=e^{-{it\over hbar}H}$. In fact, many of the superpositions mentioned as examples in Section \ref{sec:intro} can be shown to be of this form for a specific time $t$. Because the superposition size measure of Eq.(\ref{eqn:disting}) depends on the maximal probability $p_{\text{H},\text{succ}}$ of distinguishing the branches of the superposition (considered as pure states themselves), it is useful to know how long it takes for a state to time evolve to a state from which it can be probabilistically distinguished, within a prescribed precision. 

\textit{Theorem 1:} Given $\delta \in (0,1/2)$ and $U(t):=e^{-i{H\over \hbar}t}$ with $H=H^{\dagger}$, if a $t=0$ 
state $\rho$ is optimally distinguishable from the 
time evolved state $U(t)\rho U^{\dagger}(t)$ ($t>0$) with probability $1-\delta$, then \begin{equation}
t \ge { 2\text{Sin}^{-1}(1-2\delta)\over \sqrt{\mathcal{F}(\rho,H) }} =: \tau_{\text{dist}}
\label{eqn:disttime1}\end{equation} with $\Delta H := H - \langle \phi \vert H \vert \phi \rangle$ and $\mathcal{F}(\rho,H)$ 
the quantum Fisher information of $\rho$ on the unitary path generated by $H$.

Since for pure states, e.g., $\rho = \ket{\phi}\bra{\phi}$, the quantum Fisher information $\mathcal{F}(\rho,H) = {4 \over \hbar^{2}}\langle (\Delta H)^{2} \rangle_{\ket{\phi} } $, we have the following corollary:

\textit{Corollary 1:} Let the $t=0$ pure state in Theorem 1 be $\ket{\phi}\bra{\phi}$. If $\ket{\phi}\bra{\phi}$ is optimally distinguishable from the pure state $U(t)\ket{\phi}\bra{\phi}U^{\dagger}(t)$ ($t>0$) with probability $1-\delta$, then \begin{equation}
t \ge {\hbar \text{Sin}^{-1}(1-2\delta)\over \sqrt{\langle (\Delta H)^{2} \rangle_{\ket{\phi} } }} =: \tau_{\text{dist}}
\label{eqn:disttime2}\end{equation} with $\Delta H := H - \langle \phi \vert H \vert \phi \rangle$.
Proofs of Theorem 1 and Corollary 1 are given in Appendix \ref{sec:app}.

Theorem 1 defines the distinguishability time $\tau_{\text{dist}}$ which is the minimal time to optimally distinguish (to precision $\delta$) a 
general state from its unitarily time evolved image using unrestricted measurements.
We note that 
when applied to a pure state, if the state and its time-evolved image are required to be completely distinguishable ($\delta=0$), then Eq.(\ref{eqn:disttime2}) becomes the well known Mandelstam-Tamm inequality for the orthogonalization time \cite{mandelstamtamm}, namely, $t\ge {\pi \hbar \over 2\sqrt{\langle (\Delta H)^{2} \rangle_{\ket{\phi} } }}$. For $0 < \delta < 1/2$, the $\delta$-dependent distinguishability time $\tau_{\text{dist}}$ is strictly less than this orthogonalization time.

We can also use the variance of the generator of evolution $H$ to bound the rate of change of $C_{\delta}(\ket{\Psi})$ through the following inequality: \begin{equation}
{\sqrt{\log (4\delta - 4\delta^{2})} \over -\sqrt{N}}{d\sqrt{C_{\delta}} \over dt}\Big\vert_{t=0} \le {\sqrt{\langle (\Delta H)^{2} \rangle_{\ket{\phi}}}\over \hbar}.
\label{eqn:rate}\end{equation}
The proof of this inequality is given in Appendix \ref{sec:rate}.

According to the Margolus-Levitin inequality \cite{margolus}, if the expected energy content $\langle H \rangle$ of a pure state
 is less than the energy variance in the same state, then the orthogonalization time is bounded below more tightly than in the Mandelstam-Tamm inequality. Hence, one can define another distinguishability time $\tau_{\text{dist},\text{ML}}$ which depends on the total energy
rather than on the variance of the Hamiltonian. $\tau_{\text{dist},\text{ML}}$ is the appropriate distinguishability time to use if and only if the Margolus-Levitin orthogonalization time gives the shortest time to evolution to an orthogonal state (given that the state actually evolves to an orthogonal state). We can similarly generalize this inequality to bound the evolution time of a general (pure or mixed) state to a non-orthogonal state, with the following theorem.

\textit{Theorem 2:} Let $\delta \in (0,1/2)$ and $U(t):=e^{-i{H\over \hbar}t}$ with $H=H^{\dagger}$ a bounded observable. If the $t=0$ quantum state $\rho$ on a Hilbert space $\mathcal{H}_{A}$ is optimally distinguishable from the  
state $\rho(t) = U(t)\rho U^{\dagger}(t)$ ($t>0$) with probability $1-\delta$, then \begin{equation}
t \ge {\pi\hbar (1-\sqrt{1-(1-2\delta)^{2}})\over 2\text{min}\langle H \otimes \mathbb{I}_{B} \rangle_{\ket{\psi} } } =: \tau_{\text{dist},\text{ML}} 
\label{eqn:disttimeml}\end{equation} where the minimization is over all pure states $ \ket{\psi} \in \mathcal{H}_{A} \otimes \mathcal{H}_{B}$ such that $\rho = \text{tr}_{B}\ket{\psi}\bra{\psi}$, i.e., such that $\ket{\psi}$ is a purification of $\rho$.

This theorem is also proven in Appendix \ref{sec:app}.
It is interesting that this derivation of a Margolus-Levitin type distinguishability time produces, in general,
 a larger lower bound for the orthogonalization time than $\pi \hbar \over 2 \langle H \rangle_{\rho}$, 
which is the lower bound that one would naively 
expect from the Margolus-Levitin orthogonalization time for initial pure states. See Ref.\cite{heydari} and references therein for other approaches toward generalized the Margolus-Levitin quantum speed limit to mixed states.

It is also instructive to consider the superposition size of a state of the form $(U_{1}(t)^{\otimes N}+U_{2}(t)^{\otimes N}) \ket{\phi}^{\otimes N}$. These are superpositions of states along two different time-evolution paths in $\mathcal{H}^{\otimes N}$. Because the trace norm is unitarily invariant, this superposition exhibits the same branch distinguishability  
superposition size as the general state $\ket{\Psi}$ with $U=U_{1}^{\dagger}(t)U_{2}(t)$. In this case, the value of $\vert z \vert^{2}$ is simply the Loschmidt echo
\cite{jalabert2001environment} of $\ket{\phi}$ when the forward evolution is given by $U_{1}$ and backward evolution is given by $U_{2}$. This situation also appears when one wants to construct states which have the same form as $\ket{\Psi}$ except for 
the addition of a relative phase between the two branches. To this end, let $U_{1(2)}=\exp({-iH_{1(2)}t\over \hbar})$ and $H_{1}\ket{\phi} = \lambda_{1}\ket{\phi}$. Then $(U_{1}(t)^{\otimes N}+U_{2}(t)^{\otimes N}) \ket{\phi}^{\otimes N}$ is equal to $(\mathbb{I} + e^{iN\lambda_{1}t\over \hbar}U_{2}(t)^{\otimes N})\ket{\phi}^{N}$.

A physically important corollary of this observation is that if the unitary $U_{1}(t)$ is generated by observable $H_{1}$ and $U_{2}(t)$ is generated by observable $H_{2}$ with $H_{1} \neq H_{2}$,  
and $[H_{1},H_{2}] = 0$, then the branch distinguishability superposition size for the state $(U_{1}(t)^{\otimes N}+U_{2}(t)^{\otimes N}) \ket{\phi}^{\otimes N}$ is the same as 
that for $\ket{\Psi}$ with $U = e^{-i(H_{2}-H_{1})t / \hbar}$.

By considering the general case of distinguishing two quantum states time-evolving according to different paths by $U_{1}(t)$ and $U_{2}(t)$, one is led to important bounds on the time derivative of the maximal success probability of distinguishing two independently evolving quantum states. Making use of Eq.(\ref{eqn:frowis}) and the reverse triangle inequality ($\vert \Vert x \Vert - \Vert y \Vert \vert \le \Vert x-y \Vert$ for $x$ and $y$ in some normed vector space) for the trace norm, it is easy to see that if $\rho_{A(B)}(t) = e^{-iH_{A(B)}t\over \hbar}\rho_{A(B)} e^{iH_{A(B)}t\over \hbar}$ are the unitarily time-evolved quantum states $\rho_{A}$ ($\rho_{B}$), then \begin{widetext}
\begin{equation}
\big\vert \Vert \rho_{A} - \rho_{B} \Vert_{1} - \Vert \rho_{A}(t) - \rho_{B}(t) \Vert_{1} \big\vert \le 2\sum_{k\in \lbrace A , B \rbrace }\sin\left( {\sqrt{\mathcal{F}(\rho_{k},H_{k})}t\over  2}\right), 
\label{eqn:reversetriang}\end{equation} 
in which we have used $\mathcal{F}(\rho,A)$ to denote the quantum Fisher information of the state $\rho$ on a path generated by  time-independent 
observable $A$ \cite{caves}. For $\rho$ a pure state, $\mathcal{F}(\rho,A) = { 4\text{tr}(\rho(\Delta A)^{2} ) \over \hbar^{2} } $ \cite{cavesmilburn}. 

We will now use Eq.(\ref{eqn:reversetriang}) to derive an intriguing inequality which provides an upper bound for time derivative of the  optimal probability for successfully distinguishing two independently-evolving quantum states
that are given at time $t=0$ by $\rho_{A}$ and $\rho_{B}$,
respectively. First assume that $\rho_{A} \neq \rho_{B}$ and that $\mathcal{F}(\rho_{A},H_{A}) > \mathcal{F}(\rho_{B},H_{B})$,
 so that Eq.(\ref{eqn:reversetriang}) makes sense for $t\in [0,{2\pi \over \sqrt{\mathcal{F}(\rho_{A},H_{A})} - \sqrt{\mathcal{F}(\rho_{B},H_{B})}}]$ (which results in a positive semidefinite right hand side of the inequality). From Eq.(\ref{eqn:reversetriang}) and the definition of $p_{\text{H},\text{succ}}$ in Section \ref{sec:size} 
it follows that: \begin{eqnarray} \big\vert \left( {dp_{\text{H},\text{succ}}(\rho_{A}(t),\rho_{B}(t))\over dt } \right)_{t=0} \big\vert & = &\lim_{t\rightarrow 0} \big\vert {\Vert \rho_{A}(t) - \rho_{B}(t) \Vert_{1} - \Vert \rho_{A} - \rho_{B} \Vert_{1} \over 4t} \big\vert \nonumber \\ & \le &{1\over 4}\left(\sqrt{\mathcal{F}(\rho_{A},H_{A})} + \sqrt{\mathcal{F}(\rho_{B},H_{B})} \right) \nonumber \\ {}&\le & {1\over \hbar}{ \sqrt{\text{tr}(\rho_{A}(\Delta H_{A})^{2})} + \sqrt{\text{tr}(\rho_{B}(\Delta H_{B})^{2})}\over 2},
\label{eqn:derivative}\end{eqnarray} \end{widetext}
with the second inequality becoming an equality 
when both $\rho_{A}$ and $\rho_{B}$ are pure states. In going from first to the second line in Eq.(\ref{eqn:derivative}), we have used that $\lim_{x\rightarrow 0} \sin(ax)/x = a$ for $a\in \mathbb{R}$.

From Theorem 1, Eq.(\ref{eqn:derivative}), and Eq.(\ref{eqn:rate}), it is clear that there is a close relationship between the binary distinguishability of pure states on a path $\lbrace \ket{\phi(t)} = U(t)\ket{\phi} = e^{-iHt\over \hbar}\ket{\phi} \vert 0 \le t\le T \rbrace$ in Hilbert space and the Fubini-Study distance $ds_{\text{FS}}^{2} = {4 \langle (\Delta H)^{2}\rangle_{\ket{\phi}} \over \hbar^{2}} dt^{2}$ \cite{bengt,caves} on this path. This relationship can be established by noting that $\vert \langle \phi \vert U(dt) \vert \phi \rangle \vert^{2} = 1- {\langle (\Delta H)^{2}\rangle_{\ket{\phi}} \over \hbar^{2}} dt^{2} + O(dt^{3})$ \cite{anandan}, and using Eq.(\ref{eqn:rdmprob}) with $n=1$ to write the Fubini-Study line element on the projective single-mode Hilbert space (i.e., the single-mode Hilbert space modulo the equivalence relation identifying vectors which are complex scalar multiples of each other) in terms of the maximal probability of distinguishing infinitesimally separated states on a path: \begin{equation}
 ds_{\text{FS}}^{2} = (2p_{\text{H},\text{succ}}(\ket{\phi}\bra{\phi} , U(dt)\ket{\phi}\bra{\phi}U^{\dagger}(dt)) - 1)^{2}.
\label{eqn:fubinistudynew}
\end{equation}

\section{\label{sec:variance}Metrological macroscopicity and distinguishability times in spin systems}
Recently, a measure of the effective size of a quantum state on $(\mathbb{C}^{2})^{\otimes N}$ was introduced in \cite{dur} with the intention of defining macroscopicity of a quantum state by the existence of an extensive scaling ($\propto N^{-1}$ with $N$ the number of modes) of the maximum achievable precision of an estimator of a parameter defining the quantum state. This type of scaling with the number of modes in a single-shot experiment is usually called the ``Heisenberg limit'' \cite{giovannetti2011advances,zwierz2012ultimate}
and is a factor of ${1\over \sqrt{N}}$ lower than the ``standard quantum limit'' attainable from a product state of $N$ modes.
The ${1\over N}$ scaling allows the state to be considered ``metrologically useful.''
The proposed measure of effective size of a quantum state $\rho$ of a spin-1/2 system is given
in Ref.\cite{dur}  by: \begin{equation} N^{F}(\rho) : = \text{max}_{A=A^{\dagger}}{\hbar^{2}\mathcal{F}(\rho,A)\over 4N} \label{eqn:qfimacro}
\end{equation} where the maximization is carried out over $1$-local observables ($A = \sum_{i=1}^{N}A^{(i)} \otimes \mathbb{I}^{N\setminus \lbrace i \rbrace}$ with $\Vert A^{(i)} \Vert =1$) \cite{dur}. 
Note that if $\rho$ is a pure state, then the quantity maximized over in the definition of $N^{F}$ is ${\text{tr}(\rho (\Delta A )^{2} ) \over N}$. 
A quantum state $\rho$ is then considered macroscopic if it satisfies $N^{F}(\rho) \in \mathcal{O}(N)$. 
$N^{F}$ was further used to derive a measure of superposition size for states of the form $\ket{\psi}\propto \ket{A} + \ket{B}$ by taking the ratio of the value of $N^{F}$ in $\ket{\psi}$ to the average of the $N^{F}$ values of $\ket{A}$ and $\ket{B}$: 
\begin{equation}
N^{rF}(\ket{\psi}) = {N^{F}(\ket{\psi}\bra{\psi}) \over {1\over 2}  \left( N^{F}(\ket{A}\bra{A}) + N^{F}(\ket{B}\bra{B}) \right) }.
\label{eqn:relfisher}\end{equation}
$N^{rF}$ is defined as the ``relative Fisher information'' of the superposition state: such a superposition state is then referred to as  macroscopic when $N^{F}(\rho) \in \mathcal{O}(N)$.

In quantum metrology, the quantum Fisher information $\mathcal{F}(\rho,A)$, determines the maximal achievable precision of an estimator of the evolution parameter $\theta$ for unitary evolution of a state $\rho$ under $U =e^{i\theta A}$, through the quantum Cram\'{e}r-Rao bound \cite{caves}. In this metrological context, the definition Eq.(\ref{eqn:relfisher}) clearly implies that if an $N$-mode, two branch superposition probe 
state exhibits $N^{rF} \in \mathcal{O}(N)$, then there exists a 1-local Hamiltonian $A$ for which the superposition can be used to estimate the phase $\theta$ on the unitary path generated by $A$ to $\mathcal{O}(1/N)$ precision (assuming a single-shot experiment). When the individual branch states of such a superposition are used as probes, the precision in estimation of $\theta$ is limited by $\mathcal{O}(1/\sqrt{N})$ regardless of the 1-local Hamiltonian generating the evolution.  

We now state and prove a theorem which is useful for relating the superposition size measures in Eq.(\ref{eqn:disting}) and Eq.(\ref{eqn:relfisher}) for the case of equal-amplitude superposition states of the general form $\ket{\Psi}$
in finite spin-1/2 systems, i.e, when the single mode Hilbert space is $\mathcal{H} \equiv (\mathbb{C}^{2})^{\otimes N}$.

\textit{Theorem 3:} Given a state of the form 
$\ket{\Psi}\in (\mathbb{C}^{2})^{\otimes N}$ with $U \neq \mathbb{I}$, let $P_{+}$ ($P_{-}$) be the projector onto the subspace of the single-mode Hilbert space $\mathcal{H}$ associated with the positive (negative) eigenvalue of the following difference of rank-1 projectors: $\ket{\phi}\bra{\phi} - U\ket{\phi}\bra{\phi}U^{\dagger}$. Then the 1-local 
self-adjoint operator $A = \sum_{i}A^{(i)}$ with $A^{(i)} = P_{+}-P_{-}$ for all $i$ satisfies: \begin{equation}
\langle (\Delta A)^{2} \rangle_{\ket{\Psi}} = {N^{2}(1-\vert z \vert^{2}) + N(\vert z\vert^{2} + \text{Re}z) \over 1+\text{Re}(z^{N})},
\label{eqn:variance}\end{equation}
where $z$ was defined previously (Eq.~(\ref{eqn:state})).

Proof of this theorem is found in Appendix \ref{sec:variancethm}. An important feature of this result is that as $\vert z\vert$ is decreased from $1$ to $0$, the variance of this 1-local observable 
in the state $\ket{\Psi}$ increases from $N$ to $N^{2}$. 
As a consequence, given such a superposition state $\ket{\Psi} \in \mathbb{C}^{\otimes N}$ and a system Hamiltonian defined as the 1-local self-adjoint operator $A$ appearing in Theorem 3, the phase $\theta$ along the unitary path that the Hamiltonian generates can be estimated with precision limited by the standard quantum limit 
for $\vert z \vert \approx 1$ and by the Heisenberg limit for $\vert z \vert \ll 1$.

For spin-1/2 systems, the single particle Hilbert space $\mathcal{H}=\mathbb{C}^{2}$, and the maximal variance in 
each of the branches of $\ket{\Psi}$ for the set of 1-local observables with unit operator norm scales linearly with $N$ \cite{smerzi}. Then it is evident that as long as the overlap $z$ has small magnitude, the superposition $\ket{\Psi}$ is ``metrologically macroscopic'' according to the measure $N^{rF}$, i.e., $N^{rF}\in \mathcal{O}(N)$.
Since we have shown above that for $\ket{\Psi} \in (\mathbb{C}^{2})^{\otimes N}$ the branch distinguishability size $C_{\delta}(\ket{\Psi})$ is also $\mathcal{O}(N)$, we may conclude that the two measures $C_{\delta}(\ket{\Psi})$ and $N^{rF}$ give the same linear superposition size scaling in the number of modes $N$ of $\ket{\Psi}$.

Note that a POVM containing $P_{+}$ and $P_{-}$ is sufficient for achieving the maximal probability over all measurements of successfully distinguishing the states $\ket{\phi}\bra{\phi}$ and $U\ket{\phi}\bra{\phi}U^{\dagger}$ (i.e., the branches of the superposition $\propto \ket{\phi} + U\ket{\phi}$) \cite{fuchs}. It is intriguing that for $\vert z \vert \ll 1$, one may then use this POVM to construct (via Theorem 3) a 1-local observable which allows the superposition to exhibit large value of $N^{rF}$. We shall make further use of this construction for infinite dimensional Hilbert spaces in Section~\ref{sec:photons} below.

It must be emphasized that the 
$\mathcal{O}(N^{2})$ scaling of the superposition size measures $N^{rF}$ and $C_{\delta}$ with the number of modes for small branch overlap $\vert z\vert $ demonstrated here is restricted to states of the form $\ket{\Psi}$. In general, there can exist superpositions exhibiting $\mathcal{O}(N)$ value of $C_{\delta}$ while simultaneously exhibiting a microscopic $\mathcal{O}(1)$ value of $N^{rF}$ \cite{dur,frowislink}. This happens when one or both of the branches in the superposition exhibit large $\mathcal{O}(N^{2})$ maximal variance but remain distinguishable (with high maximal probability) by measurements of a small ($\mathcal{O}(1)$) number of modes.

Because the variance of the generator of unitary time evolution appears in Eq.(\ref{eqn:disttime2}), that inequality can now be used to reformulate the definition of $N^{rF}$ to take into account the different pure state dynamics of a superposition and its branches under unitary time evolution. For a pure state $\ket{\psi}$ that evolves in time according to $U(t)=e^{-i{t\over \hbar}H}$, we define a distinguishability time $\tau_{\text{dist}}(\ket{\psi},\delta ,H) :={ \hbar \text{Sin}^{-1}(1-2\delta) \over \sqrt{\langle (\Delta H)^{2} \rangle_{\ket{\psi}}}} $  (equal to the lower bound in Eq.(\ref{eqn:disttime2})) representing the minimal time required for the state $\ket{\psi}$ to time evolve to another pure state from which it is optimally distinguishable with probability $1-\delta$.
Now the maximal variance over 1-local observables in any state is the same as that in state after rotation by a local unitary. To see this, first let $V = \otimes_{j=1}^{N}U_{i}$ with all the $U_{i}$ unitary then note that if $H$ is 1-local, $V^{\dagger}HV$ is clearly 1-local. This fact implies that the branches of $\ket{\Psi}$ exhibit the same maximal value of the variance over all 1-local observables.
Hence, for all $\delta \in (0,1/2]$, we have the following expression for $N^{rF}(\ket{\Psi})$:
\begin{equation}
N^{rF}(\ket{\Psi}) = \left( {\text{min}_{H}\tau_{\text{dist}}(\ket{\Phi},\delta ,H) \over \text{min}_{H} \tau_{\text{dist}}(\ket{\Psi},\delta ,H)} \right)^{2},
\label{eqn:timeratio}\end{equation} where $\ket{\Psi}$ and $\ket{\Phi}$ are as in Eq.~(\ref{eqn:state}) and 
we have made use of the equivalence between minimization over distinguishability time and maximization over variance of $H$. The minimizations in numerator and denominator are carried out over the same set of observables as the maximizations appearing in the definition of $N^{rF}$, Eq.(\ref{eqn:qfimacro}).

\section{\label{sec:photons}Metrological macroscopicity of photonic superpositions}
In the discussion of Section \ref{sec:variance}, it was seen that if a two branch superposition of an $N$-mode spin system exhibits $N^{rF} \in \mathcal{O}(N)$ then a single-shot quantum Cram\'{e}r-Rao bound of $\mathcal{O}(1/ N)$ for parameter estimation can be achieved by the superposition, but not by its branches. This fact carries over to photonic systems in a 
much more dramatic form. Specifically, it was shown in Ref.\cite{volkoff}, that if the restriction of unit operator norm in the maximization in the definition $N^{rF}$ is removed by allowing the inclusion of certain unbounded observables in the maximization, there exist equal amplitude, two branch, $N$-mode superpositions of photonic states with $N^{rF} \in \mathcal{O}(N\langle a^{\dagger} a \rangle)$ where $\langle a^{\dagger} a \rangle$ is the expected number of photons in the state. This improvement implies that for such states, the quantum Cram\'{e}r-Rao bound is given by the total number of photons instead of just the number of modes (one could call this ``sub-Heisenberg'' limited precision). However, given a two-branch photonic superposition, a general construction of the appropriate algebra of observables to use in the definition of $N^{rF}$ in order to give such extensive scaling was lacking until now.

In view of Theorem 3, which is formulated 
explicitly for spin systems, one may ask if an observable $O$ (in an algebra $\mathfrak{a}$) with variance $\in \mathcal{O}(N^{2}\langle a^{\dagger} a \rangle^{k})$, with $k\ge 1$, may be found for a photonic state having the form $\ket{\Psi}$ in the similar way that an observable with variance $\in \mathcal{O}(N^{2})$ was constructed for a spin system superposition $\ket{\Psi}$ having small branch overlap 
$\vert z\vert$ (see Eq.(\ref{eqn:variance})). If this is possible, then $N^{rF} \in \mathcal{O}(N \langle a^{\dagger} a \rangle)$ 
when the maximization is restricted to the algebra $\mathfrak{a}$ such that 
$O \in \mathfrak{a} $ and \begin{equation} \text{max}_{A \in \mathfrak{a}}\langle (\Delta A)^{2}\rangle_{\ket{\Phi}} +\text{max}_{A \in \mathfrak{a}} \langle (\Delta A)^{2}\rangle_{V\ket{\Phi}} \in \mathcal{O}(N\langle a^{\dagger}a\rangle^{k-1}),
\end{equation}where $A$ is a 1-local self-adjoint operator. Here we illustrate a procedure 
to find the appropriate algebra $\mathfrak{a}$ for entangled coherent states, 
multi-mode Fock states, and the hierarchical cat states.
Identification of such algebras of 1-local observables with allowing for superposition sizes $\ket{\Psi}$ to exhibit $N^{rF} \in \mathcal{O}(N \langle a^{\dagger} a \rangle)$  has significant metrological implications, as we discuss at the end of this section.

\subsection{Entangled coherent states\label{subsec:ecs}}

We consider here $N$-mode entangled coherent states and restrict our attention to the even states $\ket{\alpha}^{\otimes N} + \ket{-\alpha}^{\otimes N}$. The usefulness of this state in quantum metrology protocols has been noted previously \cite{infoprocesscat}. As noted above these may be written as
\begin{equation} 
\ket{\text{ECS}_{N}(\alpha)}  \propto  (\mathbb{I} + \otimes_{k=0}^{N}e^{-i\pi a_{k}^{\dagger}a_{k}})\ket{\alpha}^{\otimes N} 
\end{equation}
with normalization constant $1/\sqrt{2+2e^{-2N\vert \alpha \vert^{2}}}$.  
When 1-local quadrature operators 
$x^{(\theta)} := {1\over \sqrt{2}}(ae^{-i\theta} + a^{\dagger}e^{i\theta})$
are included in the maximization defining $N^{rF}$,  we claim that $N^{rF}$ satisfies 
$N^{rF}\ge N\vert \alpha \vert^{2}\tanh N\vert \alpha \vert^{2} + \vert \alpha \vert^{2} + {1\over 2N}$, i.e., 
the superposition size scales $\in \mathcal{O}(N \langle a^{\dagger} a \rangle)$.
We now proceed to prove this claim in a way which allows for generalization to many two-branch photonic superpositions. 
 
To this end, consider an orthonormal basis $\lbrace \ket{\psi_{+}},  \ket{\psi_{-}} \rbrace$ which defines a 2-D subspace $\mathcal{K}$ of $\ell^{2}(\mathbb{C})$, where $\ket{\psi_{\pm}} \propto \ket{\alpha} \pm \ket{-\alpha}$ are the even ($+$) and odd ($-$) coherent states, respectively \cite{dodonov}. It is then clear that $\ket{\alpha}\bra{\alpha} - \ket{-\alpha}\bra{-\alpha} = \sqrt{1-e^{-4\vert \alpha \vert^{2}}}\sigma_{x}$, with $\sigma_{x}$ the appropriate Pauli matrix in the subspace defined by $\lbrace \ket{\psi_{+}}, \ket{\psi_{-}} \rbrace$. 
According to Theorem 3, and the fact that $\langle \alpha \vert -\alpha \rangle = e^{-2\vert \alpha \vert^{2}}$, the operator $\sum_{i=1}^{N}P_{+}^{(i)} - P_{-}^{(i)}$ (where $P_{\pm} = (1/2)(\ket{\psi_{+}} \pm \ket{\psi_{-}})(\bra{\psi_{+}} \pm \bra{\psi_{-}})$ are the projectors onto the orthogonal eigenspaces of  $\sqrt{1-e^{-4\vert \alpha \vert^{2}}}\sigma_{x}$ in the given basis of $\mathcal{K}$,
so that the spectral decomposition of $\sigma_{x}$ is $ \sigma_{x}=P_{+} - P_{-}$) is a 1-local observable having variance in $\mathcal{O}(N^{2})$ in the entangled coherent state. In the individual branches of $\ket{\text{ECS}_{N}(\alpha)}$, the maximal variance over the spin-1/2 observables of the Hilbert subspace $\mathcal{K}$ is $N$,
because they are product states.

Since the $N$-mode electric fields corresponding to the branches of $\ket{\text{ECS}_{N}(\alpha)}$ are $\pi$ out of phase, we expect that a quadrature operator should exhibit a large variance in $\ket{\text{ECS}_{N}(\alpha)}$ but a small variance in each of its branches. We now note that in the subspace $\mathcal{K}$, the operator $\sigma_{x}$ is weakly equivalent to such an unbounded operator (i.e., has the same expectation value in all states of $\mathcal{K}$), as follows:

\begin{eqnarray} \sigma_{x} &\sim & {1\over 2\vert \alpha \vert}(ae^{-i\text{Arg}(\alpha)}+a^{\dagger}e^{i\text{Arg}(\alpha)})\sqrt{1 - e^{-4\vert \alpha \vert^{2}}} \nonumber \\ &{=}& { \sqrt{{1\over 2} - {1\over 2}e^{-4\vert \alpha \vert^{2}}}\over \vert \alpha  \vert} x^{( \text{Arg}(\alpha))} \label{eqn:ecssim} \end{eqnarray} 
Here we have used the quadrature operator $x^{(\theta)}$.
Writing $P_{\mathcal{K}} = P_{+} + P_{-}$ as the projection onto the subspace $\mathcal{K}$, we have the equality $\sigma_{x} =cP_{\mathcal{K}}x^{( \text{Arg}(\alpha))}P_{\mathcal{K}}$, with $c = \sqrt{{1\over 2} - {1\over 2}e^{-4\vert \alpha \vert^{2}}} / \vert \alpha  \vert$. It is clear that if one can write $P_{+}-P_{-}=cP_{\mathcal{K}}OP_{\mathcal{K}}$ for an unbounded observable $O=O^{\dagger}$ and $c\in \mathbb{R}$, then for a state of the form $\ket{\Psi}$ in $ (\ell^{2}(\mathbb{C}))^{\otimes N}$, the operator $A$ of Theorem 3 has at least ${1\over  c^{2}}$ times the variance when $A^{(i)} = O$ as when $A^{(i)} = P_{+} - P_{-}$. Specifically, this is a consequence of the fact that for any $\ket{ \varphi } \in \mathcal{K} \subset \ell^{2}(\mathbb{C})$, one has $\langle \varphi \vert O \vert {\varphi}\rangle={1\over c}\langle \varphi \vert P_{+}-P_{-} \vert \varphi \rangle$ while at the same time \begin{eqnarray}\langle \varphi \vert O^{2} \vert \varphi \rangle &-&{1\over c^{2}}\langle \varphi \vert (P_{+}-P_{-})^{2} \vert \varphi \rangle  \nonumber \\ &=& \langle \varphi \vert O(\mathbb{I}_{\mathcal{H}}-P_{\mathcal{K}})O \vert \varphi \rangle  > 0 \end{eqnarray} because $ \mathbb{I}_{\mathcal{H}} - P_{\mathcal{K}} >0$. Hence, $\langle (\Delta O)^{2} \rangle_{\ket{\varphi}} \ge {1\over c^{2}}\langle (\Delta(P_{+} - P_{-}))^{2}\rangle_{\ket{\varphi}}$. In the present case, Eq.(\ref{eqn:ecssim}) shows that $1/c^{2} \in \mathcal{O}(\vert \alpha \vert^{2})$.
Hence, the maximal variance of 1-local quadrature operators in the entangled coherent state is in $\mathcal{O}(N^{2}\vert \alpha\vert^{2})$.
 
As discussed in~\cite{volkoff}, the variance of 1-local 
quadrature operators scales with the number of modes as $\mathcal{O}(N)$ in each of the branches $\ket{\alpha}^{\otimes N}$ and $\ket{-\alpha}^{\otimes N}$ of $\ket{\text{ECS}_{N}(\alpha)}$, since these are product states. 
The lack of dependence on photon number of the variance of the 1-local quadrature operator in the branches may also be viewed as resulting from the fact that all quadratures have the same variance in a coherent state and that it is an intelligent state for the Heisenberg uncertainty relation \cite{brif}, i.e., it saturates the lower bound in the Heisenberg uncertainty relation.
The individual branches will then have values of $N^{F}$ which are ``microscopic,'' 
i.e., in $\mathcal{O}(1)$.  This feature of the individual branches combined with the 
fact that the maximal variance in the entangled coherent state is in $\mathcal{O}(N^{2}\vert \alpha\vert^{2})$
when quadrature operators (i.e., a representation on $\ell^{2}(\mathbb{C})$ of the complexification of the Heisenberg algebra $\mathfrak{h}_{3}$) are included in the maximization over 1-local observables, yields 
a value of $N^{rF} \in \mathcal{O}(N\langle a^{\dagger}a \rangle )$ \cite{volkoff}.

We note further that if we were to extend the optimization in the definition of $N^{rF}$ to the 1-local observables derived from the oscillator Lie algebra $\mathfrak{h}_{4} = (\text{span}\lbrace a,a^{\dagger},a^{\dagger}a,\mathbb{I} \rbrace , [\cdot , \cdot])$, the entangled coherent state $\ket{\text{ECS}_{N}(\alpha)}$ would now have an $N^{rF}$ value of  only $\mathcal{O}(N)$, due to the fact that in the product state branch $\ket{\alpha}^{\otimes N}$, the observable $A = \sum_{i=1}^{N}A^{(i)}$ with $A^{(i)} = a^{\dagger}a$ for all $i$, has a variance scaling linearly with the number of photons, $\vert \alpha \vert^{2}$. Hence, $\mathfrak{h}_{3}$ is a ``minimal'' algebra allowing for 1-local observables to exhibit a maximal variance in $\mathcal{O}(N\langle a^{\dagger}a\rangle)$ for the entangled coherent state.

\subsection{Superpositions of multi-mode Fock states}

There are other quantum superpositions of the form of $\ket{\Psi}$, for which $\mathfrak{h}_{4}$ actually is the smallest algebra containing an observable allowing for a value of $N^{rF}$ 
that depends on the excitation number. 
We consider here GHZ-like superpositions of multi-mode Fock states
$(1/ \sqrt{2})(\ket{0}^{\otimes N} + \ket{n}^{\otimes N})$.
Note that for $N=2$, this state can be locally rotated to the well-known N00N state $(1/ \sqrt{2})(\ket{n}\ket{0} + \ket{n}\ket{0})$. 
One may write a general Hamiltonian of $\mathfrak{h}_{4}$ by $L(\ell , \beta) : = \ell a^{\dagger}a + {\overline{\beta}a + \beta a^{\dagger} \over \sqrt{2}}$ with $\ell \in \mathbb{R}$ and $\beta$ lying on the unit circle in the complex plane. Taking the frequency $\ell =1$ for simplicity, one finds that 
$(1/ \sqrt{2})(\ket{0}^{\otimes N} + \ket{n}^{\otimes N})$ has $N^{rF} = {Nn \over 4(1+{1\over n})} + {1\over 2}$ when $n > 2$ and 
when the maximization over 1-local observables is carried out over $\mathfrak{h}_{4}$. 
This result can be made to follow the example of the entangled coherent state. To this end we note the following facts:

1) By the spectral theorem for observables on $\mathcal{K}$, the Pauli operator $\sigma_{z}$ is equal to $P_{+}-P_{-} = \ket{n}\bra{n}-\ket{0}\bra{0}$, where the action of $\sigma_{z}$ in $\mathcal{K}$ is $\sigma_{z}\ket{0} = -\ket{0}$ and $\sigma_{z}\ket{n} = \ket{n}$.
Thus $\sigma_{z}$ coincides with the difference of the elements of the POVM which allows for distinguishing $\ket{0}$ from $\ket{n}$ with unit probability.  

2) With the action defined in 1), $\sum_{i=1}^{N}\sigma_{z}^{(i)}$ has variance $N^{2}$ in $(1/2)(\ket{0}^{\otimes N} + \ket{n}^{\otimes N})$ so that $N^{rF} \in \mathcal{O}(N)$ if the maximization is carried out over $\mathfrak{su}(2,\mathbb{C})$ by the same reasoning as in Theorem 3.

3) We can find an unbounded operator in 
$\mathfrak{h}_{4}$ which gives $\sigma_{z}$ when compressed down to the subspace of interest. Specifically, $P_{\mathcal{K}}(2a^{*}a  - n\mathbb{I} ) P_{\mathcal{K}} = n\sigma_{z}$ (so that the proportionality constant $c$ introduced in the last Subsection is $1/n$).

Hence the 1-local observable $\sum_{i=1}^{N}2a_{i}^{\dagger}a_{i} - n\mathbb{I}_{i}$ formed from $\mathfrak{h}_{4}$ has variance 
$\in \mathcal{O}((Nn)^{2})$ for the N00N state. Since the observables of $\mathfrak{h}_{4}$ have variance at most $\in \mathcal{O}(Nn)$ in the branches of state  $(1/ \sqrt{2})(\ket{0}^{\otimes N} + \ket{n}^{\otimes N})$, we conclude that $N^{rF}$ of this state is $\in\mathcal{O}(Nn)$ when the maximizations are carried out over $\mathfrak{h}_{4}$.

\subsection{Hierarchical cat states\label{subsec:HCS}}

We now  analyze $N^{rF}$ for the hierarchical cat state $\ket{\text{HCS}_{N}(\alpha)}$ introduced in Ref. \cite{volkoff}: \begin{equation}
\ket{\text{HCS}_{N}(\alpha)} : = {1\over \sqrt{2}}\left( \ket{\psi_{+}}^{\otimes N} + \ket{\psi_{-}}^{\otimes N} \right)
\end{equation}
It is clear that $P_{+}-P_{-}$ for this state is simply $\sigma_{z}$, where the action of $\sigma_{z}$ is defined in the 2-D Hilbert subspace $\mathcal{K}$ of $\ell^{2}(\mathbb{C})$ given by $\mathcal{K} := \text{span}\lbrace \ket{0} := \ket{\psi_{+}} , \ket{1}:= \ket{\psi_{-}} \rbrace $. Here, $\ket{\psi_{\pm}}$ are the superpositions of coherent states introduced in Subsection \ref{subsec:ecs}. In this sense, the $P_{+}-P_{-}$ operator is analogous to the corresponding operator for the GHZ state for a chain of $N$ spin-1/2 degrees of freedom. To show that $\ket{\text{HCS}_{N}(\alpha)}$ can exhibit a value of $N^{rF}$ scaling with the expected total number of photons in the state rather than simply the number of modes, $N$, we proceed as above and identify an unbounded operator which is weakly equivalent to $\sigma_{z}$ in $\mathcal{K}$ (i.e., has the same matrix elements as $\sigma_{z}$ in the subspace $\mathcal{K}$). To see this, we use the identity $e^{i\theta a^{\dagger}a}\ket{\psi_{\pm}} = \pm \ket{\psi_{\pm}}$, which implies: \begin{equation}
\sigma_{z} = {1\over 2\text{Re}(\alpha^{2})}P_{\mathcal{K}}(e^{i\pi {a^{\dagger}}a}a^{2} + a^{\dagger 2}e^{-i\pi a^{\dagger}a })P_{\mathcal{K}}.
\end{equation} In other words, when projected to act on the subspace $\mathcal{K}$, the operator $e^{i\pi {a^{\dagger}}a}a^{2} + a^{\dagger 2}e^{-i\pi a^{\dagger}a }$ is equal to ${ 2\text{Re}(\alpha^{2})}\sigma_{z}$. Because $\sigma_{z}=P_{+}-P_{-}$, the variance of $2\text{Re}(\alpha^{2})\sum_{i=1}^{N}\sigma_{z}^{(i)}$ in $\ket{\text{HCS}_{N}(\alpha)}$ is in $\mathcal{O}(N^{2}\vert \alpha \vert^{4})$. 

We now identify a minimal algebra of observables which allows for the ratio of the maximal variance (over 1-local Hamiltonians constructed from elements of the algebra) in $ \ket{\text{HCS}_{N}(\alpha)}$ to the maximal variance in $\ket{\psi_{\pm}}^{\otimes N}$ to be $\mathcal{O}(N\vert \alpha \vert^{2})$. To this end, note the following commutation relations:
\begin{align}
[e^{i\pi {a^{\dagger}}a}a^{2} , a^{\dagger 2}e^{-i\pi a^{\dagger}a } ] &= 4a^{\dagger}a+2 \\
[a^{\dagger}a , e^{i\pi {a^{\dagger}}a}a^{2} ] &= -2 e^{i\pi{a^{\dagger}}a}a^{2} \\
[a^{\dagger}a , a^{\dagger 2}e^{-i\pi a^{\dagger}a }] &= 2a^{\dagger 2}e^{-i\pi a^{\dagger}a } .
\end{align}

Hence, performing the rescaling $e^{i\pi {a^{\dagger}}a} \mapsto {1\over 2}e^{i\pi {a^{\dagger}}a}$, we reproduce the Lie algebra $\mathfrak{sl}(2,\mathbb{C}):= (\text{span}\lbrace  1/4(2a^{\dagger}a +\mathbb{I}), a^{2}/2 , a^{\dagger 2}/2   \rbrace , [\cdot , \cdot ] )$ except that we now use $\lbrace  1/4(2a^{\dagger}a +\mathbb{I}),e^{-i\pi a^{\dagger}a } a^{2}/2 , a^{\dagger 2}e^{-i\pi a^{\dagger}a }/2   \rbrace$ as a basis. By writing a general Hamiltonian of $\mathfrak{sl}(2,\mathbb{C})$ in this basis and calculating its variance in $\ket{\psi_{\pm}}$, it is seen that the maximal variance in the 
individual branches of $\ket{\text{HCS}_{N}(\alpha)}$ is now in $\mathcal{O}(N\vert \alpha \vert^{2})$.
We therefore conclude that $N^{rF}(\ket{\text{HCS}_{N}(\alpha)}) \in \mathcal{O}(N \vert \alpha \vert^{2})$ when the maximizations in the definition of $N^{rF}$ are carried out over 1-local Hamiltonians from $\mathfrak{sl}(2,\mathbb{C})$.

In Appendix \ref{sec:povms}, we note an important corollary of the above analyses of $N$-mode photonic states: the optimal POVMs for distinguishing the branches of photonic states of the form $\ket{\Psi}$ with maximal probability can be implemented either by directly designing the appropriate projection operators or by performing certain photonic measurements in an appropriate 2-D subspace containing $\ket{\phi}$ and $U\ket{\phi}$.

\subsection{Algebras for $N^{rF}$ macroscopicity} 
The above analysis  of three types of photonic states defined in the infinite dimensional space $\ell^{2}(\mathbb{C})^{\otimes N}$ show that the metrological size $N^{rF}$ is strongly dependent on the algebra used in its definition. This has significant implications for the optimal precision of measurements of an evolution parameter when these states are used as probes, since the Hamiltonian being estimated must be an element of the appropriate minimal algebra of operators. Conversely, given an experimentally available set of parametric Hamiltonians, one can determine what is the maximum estimation precision accessible with a given photonic superposition state.

Consider for example the entangled coherent state, $\ket{\text{ECS}_{N}(\alpha)}$. The analysis in Subsection~\ref{subsec:ecs} above shows that if
$\ket{\text{ECS}_{N}(\alpha)}$ is used as a probe in a parameter estimation protocol for an evolution operator which 
is generated by a 1-local Hamiltonian constructed from elements of $\mathfrak{h}_4$, the precision will have optimal scaling of $\mathcal{O}(1/N)$.  However, if  the algebra $\mathfrak{h}_{3}$ is used instead of $\mathfrak{h}_{4}$, then an experiment using $\ket{\text{ECS}_{N}(\alpha)}$ as probe can yield optimal parameter precision scaling in $\mathcal{O}(1/N\vert \alpha\vert^{2})$.  Hence, the Lie algebra $\mathfrak{h}_{3}$ (i.e., its complex representations as linear operators on $\ell^{2}(\mathbb{C})$) should be considered as a ``metrological macroscopicity algebra'' for $\ket{\text{ECS}_{N}(\alpha)}$.   Likewise, the two algebras $\mathfrak{h}_{4}$ and $\mathfrak{sl}(2,\mathbb{C})$ can be considered as the metrological macroscopicity algebras for ${1\over \sqrt{2}}(\ket{0}^{\otimes N} + \ket{n}^{\otimes N})$ and $\ket{\text{HCS}_{N}(\alpha)}$, respectively, because each of the algebras allows $N^{rF}$ to scale with the total number of photons for its corresponding state.
From this analysis, we conclude that a photonic superposition state $\ket{\Psi}$ may serve as a more valuable practical resource for quantum metrology than its product state branches, but this increase in utility depends on the set of Hamiltonians accessible in the experimental parameter estimation protocol, the intrinsic structure of the state, and of course, the measurements which can be made on the state.

The dependence of the metrological usefulness of $\ket{\Psi}$ on  the set of implementable Hamiltonians revealed by this analysis allows one to draw a quantitative parallel between $N^{rF}$ and the branch distinguishability measure for $\ket{\Psi}$ in $\ell^{2}(\mathbb{C})^{\otimes N}$. Recall that for spin systems, Theorem 3 gives instructions on how $N^{rF} \in \mathcal{O}(N)$ can be achieved if $C_{\delta} \in \mathcal{O}(N)$ for $\ket{\Psi}$.  In the case of photonic systems, one can consider the values of $N^{rF}$ obtained for, e.g., the superposition of multimode Fock states $(1/\sqrt{2})(\ket{0}^{\otimes N} + \ket{n}^{\otimes N})$ and the entangled coherent state $\ket{\text{ECS}_{N}(\alpha)}$, by maximization over their respective metrological macroscopicity algebras. The former state has orthogonal 
branches, so it is straightforward to apply the generalized branch distinguishability measure, Eq.~(\ref{eqn:general}), which yields $C_{\delta}$ equal to the expected total number of particles in the state, in agreement with the results of $N^{rF}$.
For $\ket{\text{ECS}_{N}(\alpha)}$, the fact that $\log \vert z \vert$ scales linearly with $\vert \alpha \vert^{2}$ produces $C_{\delta} \in \mathcal{O}(N\vert \alpha \vert^{2})$; our result for $N^{rF}$ above was seen to produce the same scaling.
Hence, the results of this Section show that $N^{rF}$ for these states scales in the same way as $C_{\delta}$. 
In general, we conjecture that if $\log \vert z \vert$ scales linearly with a given physical quantity 
$g$ encoded in $\ket{\Psi}$, a corresponding metrological macroscopicity algebra can be utilized in the definition of $N^{rF}$ to achieve scaling of $N^{rF}$ that is $N$ times that physical quantity, i.e., $Ng$.
We note that interaction induced increase of precision beyond the Heisenberg limit (for entangled states) \cite{beyondheisen} and 
beyond the standard quantum limit (for product states) \cite{beyondheisenprodstate} is one specific example of the way in which using an extended algebra of observables (e.g., by taking tensor products of $k$ copies of the algebra) can allow for greater precision in metrology protocols. 

We further note that removal of the boundedness restriction in the definition of $N^{rF}$ is not the only method to arrive at a superposition size value that is extensive in the excitation number. 
In Ref.\cite{frowislink}, a state of a single-mode radiation cavity (with Hilbert space $\ell^{2}(\mathbb{C})$) is ``imprinted'' on a spin system state (in the Hilbert space $(\mathbb{C}^{2})^{\otimes M}$ for $M$ much greater than the number of photons in the cavity) by coupling the cavity and spin system with a Jaynes-Cummings interaction. 
This attachment of a spin vacuum to the cavity state followed by rotation preserves the size of a photonic superposition state, given that the unitary mixing of a photonic state with a product state of photonic or spin vacuum cannot change the superposition size \cite{volkoff}. This procedure therefore provides a means of measuring the photon number in an arbitrary photonic state.

\section{\label{sec:conc}Conclusion}

We have analyzed two measurement-based measures of superposition size, the branch distinguishability and relative quantum Fisher information measures, for two-branch superpositions of the form $\ket{\Psi} \propto \ket{\phi}^{\otimes N} + U^{\otimes N} \ket{\phi}^{\otimes N} \sim (\mathbb{I}+V)\ket{\Phi}$ 
(Eq.~(\ref{eqn:state}), i.e., superpositions of an initial tensor product of $N$ identical single-mode states and its image under a unitary map, where the unitary is a tensor product of 
one single mode unitary. Many states of utility in quantum algorithms, quantum error correction, and quantum optics are of this form. A general formula (Eq.(\ref{eqn:general})) for the superposition size of states of the form $\ket{\Psi}$ based on binary quantum distinguishability of the branches succinctly captures the relationship between branch distinguishability and branch overlap
for these states.

When the single-mode unitary operator in $\ket{\Psi}$ is taken to be one-parameter 
time evolution, the question of branch distinguishability of $\ket{\Psi}$ becomes the question of distinguishing two pure states on the same time-evolution path in Hilbert space. We showed that the ``distinguishability time'' 
required to distinguish the two branches within a given precision, Eq.(\ref{eqn:disttime2}), provides a generalization of the orthogonalization times of Mandelstam and Tamm 
(Theorem 1) and 
of Margolus and Levitin
(Theorem 2).  In addition we have shown how the temporal rate of change of the superposition size of $\ket{\Psi}$ and the maximal probability of distinguishing its branches depend on the variance of the observable generating the path. 
It was further shown that the infinitesimal Fubini-Study line element in (projective) $\mathcal{H}$ can be written in terms of the maximal probability over unrestricted measurements of successfully distinguishing two infinitesimally displaced pure states. These results highlight the relationship between the branch distinguishability superposition size for superpositions along a time-evolution path and the statistical geometry of pure quantum states on that path.

It was shown for spin-1/2 systems (Theorem 3) that the measurement which provides the largest probability of distinguishing the branches of $\ket{\Psi}$ can be used to construct a 1-local, unit-norm observable which gives $\mathcal{O}(N^{2})$ scaling of the quantum Fisher information in $\ket{\Psi}$, as long as the branch overlap is much less than 1
(Eq. (\ref{eqn:variance})). The relative quantum Fisher information measure was further related to the distinguishability of two pure states by showing that it can be expressed as the ratio of the minimal distinguishability time of the branches of $\ket{\Psi}$ to that of the superposition itself (Eq.(\ref{eqn:timeratio})). By using the examples of $N$-mode entangled coherent states, 
superpositions of multi-mode Fock states, and hierarchical cat states, it was shown that  
for photonic systems a value of $N^{rF}$ greater than $N$ can be obtained if the maximization in the definition of $N^{rF}$ is carried out over a ``metrological macroscopicity algebra'' of observables. A general method for 
explicitly constructing this algebra in the case of quantum superpositions of the form $\ket{\Psi}$ would be a valuable tool for constructing Hamiltonians and states for metrology at the limit set by the quantum Cram\'{e}r-Rao inequality.

In terms of the logical connections among common quantum superposition macroscopicity measures \cite{frowislink}, the present work has singled out a subclass of quantum superpositions in $(\mathbb{C}^{2})^{\otimes N}$ for which macroscopicity (defined as an effective size scaling as $\mathcal{O}(N)$) according to the relative metrological usefulness measure $N^{rF}$ implies macroscopicity according to the branch distinguishability based cat size $C_{\delta}$ and \textit{vice versa}, as long as the 
value $z=\langle \phi \vert U \vert \phi \rangle$ is small. For these states, the magnitude of the inner product of the branches, $z$, provides a simple conceptual link between these two superposition size measures. In general, in view of the results of Section \ref{sec:photons}, we expect that for every two-branch superposition $\ket{\Psi}$ there exists a subalgebra of operators on $\ell^{2}(\mathbb{C})^{\otimes N}$ for which the implication (superposition macroscopicity according $N^{rF}$) $\Rightarrow$ (macroscopicity according to $C_{\delta}$) holds, although we do not currently know of a proof.

By contributing to an understanding of the logical relationships of macroscopicity measures for a class of quantum superpositions in a given Hilbert space, the results of this work will be useful for developing a structure theory of macroscopicity measures for quantum superpositions.

\appendix
\section{\label{sec:app}Proofs of 
Theorem 1, Corollary 1, and Theorem 2}
\textit{Lemma} (F. Fr\"{o}wis \cite{frowisentangle}) Given a $t=0$ quantum state $\rho$ and a quantum state $\rho(t):= e^{-iAt \over \hbar}\rho e^{iAt\over \hbar}$ with $A=A^{\dagger}$ a bounded operator, then \begin{equation}1-{1\over 4}\Vert \rho(t) - \rho \Vert_{1}^{2} \ge \cos^{2}\left( {\sqrt{\mathcal{F}(\rho,A)}t\over 2}\right) \label{eqn:frowis}\end{equation} where $\mathcal{F}(\rho,A)$ is the quantum Fisher information of the state $\rho$ on the one-parameter path generated by $A$.

\textit{Proof of 
Theorem 1}:  We rearrange Eq.(\ref{eqn:frowis}) to \begin{equation}
2\sin\left({\sqrt{\mathcal{F}(\rho,A)}t\over 2}\right) \ge \Vert \rho - \rho(t)  \Vert_{1} 
\label{eqn:step}\end{equation} Let $t$ be the time at which the maximal probability (over all possible measurements) of distinguishing two quantum states $\rho$ and $\rho(t)$ (say, with equal \textit{a priori} probabilities) is $1-\delta$. Then we have \cite{helstromone}:\begin{equation} {1\over 2}\Vert \rho - \rho(t)  \Vert_{1} = 1-2\delta . \label{eqn:trdistdelt} \end{equation}
Inserting Eq.(\ref{eqn:trdistdelt}) into Eq.(\ref{eqn:step}) gives Eq.(\ref{eqn:disttime1})
with $\text{Sin}^{-1}(x) \in (0,\pi / 2)$ for $x\in (0,1)$ being the principal branch of the inverse sine. \hspace{.5cm}$\square$

\textit{Proof of 
Corollary 1}:
Application of Theorem 1 to a pure state $\rho=\ket{\phi}\bra{\phi}$ and its unitarily time-evolved image (generated by $A=A^{\dagger}$) gives Eq.(\ref{eqn:disttime2}) because $\mathcal{F}(\ket{\phi}\bra{\phi},A) = {4 \langle (\Delta A)^{2} \rangle_{\ket{\phi}} \over \hbar^{2}}$. \hspace{.5cm}$\square$
 
Recall that $\mathcal{F}(\rho ,A) = \text{tr} \rho (L_{\rho}({d\rho \over dt})^{2}) $ where $L_{\rho }({d\rho \over dt})$ is the symmetric logarithmic derivative operator \cite{holevo}. 

\textit{Proof of 
Theorem 2}: Let $\mathcal{H}_{A}$ and $\mathcal{H}_{B}$ be Hilbert spaces with at most countably infinite dimension and let $\rho$ be a quantum state on $\mathcal{H}_{A}$. We require that $\rho$ evolves in time by a unitary generated by the bounded observable $H$. Let $\ket{\psi} \in \mathcal{H}_{A} \otimes \mathcal{H}_{B}$ be a purification of $\rho$ (so that $ \ket{\psi(t)} := e^{-{it\over \hbar}(H \otimes \mathbb{I}_{B})}\ket{\psi}$ is a purification of $\rho(t)$) which achieves the minimal value of $\langle H \otimes \mathbb{I}_{B} \rangle_{\ket{\psi}}$. According to Ref.\cite{margolus}, if $\lbrace E_{n} \rbrace$ are the eigenvalues of $H$, then for $t \in [0 , {\hbar \over \text{max}\lbrace E_{n} \rbrace} ]$, \begin{eqnarray}
\vert \langle \psi \vert \psi(t) \rangle \vert^{2} &\ge & \text{Re}\langle \psi \vert \psi(t) \rangle \nonumber \\ &\ge & \left( 1-{2\langle H \otimes \mathbb{I}_{B}\rangle_{\ket{\psi}}t\over \pi \hbar}\right)^{2}.
\end{eqnarray} We now use the following chain of inequalities: \begin{eqnarray}
1-{1\over 4}\Vert \rho - \rho(t) \Vert_{1}^{2} &\ge &\left( \text{tr}(\sqrt{\sqrt{\rho}\rho(t)\sqrt{\rho}}) \right)^{2} \nonumber \\ &=& \text{max} \vert \langle \lambda \vert \lambda(t) \rangle \vert^{2} \nonumber \\ &\ge & \vert \langle \psi \vert \psi(t) \rangle \vert^{2} \nonumber \\ & \ge & \left( 1-{2\langle H \otimes \mathbb{I}_{B}\rangle_{\ket{\psi}}t\over \pi \hbar}\right)^{2} 
\end{eqnarray} where the maximization in the second line is over all purifications of $\rho$ in $\mathcal{H}_{A}\otimes \mathcal{H}_{B}$. In the first line, we have used an inequality derived in Ref.\cite{frowisentangle} second line, and in the second line, we have used Uhlmann's theorem \cite{nielsen}. Requiring that $\rho$ and $\rho(t)$ be distinguishable with maximal success probability $1-\delta$, it follows from Eq.(\ref{eqn:trdistdelt}) that \begin{equation}
1-(1-2\delta)^{2} \ge \left( 1-{2\langle H \otimes \mathbb{I}_{B}\rangle_{\ket{\psi}}t\over \pi \hbar}\right)^{2}.
\end{equation}
Rearraging the above equation produces Eq.(\ref{eqn:disttimeml}). \hspace{.5cm}$\square$

\section{\label{sec:rate}Proof of Eq.(\ref{eqn:rate})}
The inequality follows from noting that for $\delta \in (0,{1\over 2})$: \begin{widetext} \begin{eqnarray}1+\left( {\log(4\delta - 4\delta^{2}) C_{\delta}(t) \over -N} \right) &\le & \exp({\log(4\delta - 4\delta^{2}) C_{\delta}(t) \over -N}) \nonumber \\ &=& \vert \langle \phi \vert U(t) \vert \phi \rangle\vert^{-2} \nonumber \\  &=& \left( 1 -{\langle (\Delta H)^{2} \rangle_{\ket{\phi}} \over \hbar^{2}}t^{2} + g \right)^{-1} \end{eqnarray}  where $g \in \mathcal{O}(t^{3})$. Thus, ${\log(4\delta - 4\delta^{2}) C_{\delta}(t) \over -N} \le ( {\langle (\Delta H)^{2} \rangle_{\ket{\phi}}}t^{2}/\hbar^{2} -  g )((1-{\langle (\Delta H)^{2} \rangle_{\ket{\phi}}\over \hbar^{2}}t^{2} +g)^{-1}$ and it follows that \begin{equation}
{\sqrt{\log(4\delta - 4\delta^{2}) }\sqrt{C_{\delta}(t)} \over -t\sqrt{N}} \le \sqrt{\left( {\langle (\Delta H)^{2} \rangle_{\ket{\phi}} \over \hbar^{2}} - {\vert g \vert\over t^{2}}\right) \left( 1-{\langle (\Delta H)^{2} \rangle_{\ket{\phi}} \over \hbar^{2}}t^{2} + g\right)^{-1}}
\end{equation}\end{widetext} Noting that $C_{\delta}(t=0^{+}) = 0$ for the state $\ket{\Psi}$ and that $C_{\delta}(t)$ is continuous, the limit $t\rightarrow 0$ can be taken in the above inequality, resulting in Eq.(\ref{eqn:rate}).

\section{\label{sec:variancethm}Proof of 
Theorem 3, Eq.(\ref{eqn:variance}))}:
To prove Eq.(\ref{eqn:variance}), we find the normalized eigenvectors $\ket{\xi_{+}}$ and $\ket{\xi_{-}}$ of $\ket{\phi}\bra{\phi} - U\ket{\phi}\bra{\phi}U^{\dagger}$. With $z := \langle \phi \vert U \vert \phi \rangle$, they are: \begin{widetext} \begin{eqnarray} \ket{\xi_{+}} &=&{ \vert z \vert \over \sqrt{2-2\sqrt{1-\vert z \vert^{2}}}} \left( {1\over \sqrt{1-\vert z \vert^{2}}}\ket{\phi} + {\sqrt{1-\vert z \vert^{2}} -1 \over z\sqrt{1-\vert z \vert^{2}}}U\ket{\phi} \right)\nonumber  \\ \ket{\xi_{-}} &=& { \vert z \vert \over \sqrt{2+2\sqrt{1-\vert z \vert^{2}}}} \left( {-1\over \sqrt{1-\vert z \vert^{2}}}\ket{\phi} + {\sqrt{1-\vert z \vert^{2}} +1 \over z\sqrt{1-\vert z \vert^{2}}}U\ket{\phi} \right)
\label{eqn:eigenvectors}
\end{eqnarray} \end{widetext}
corresponding to the eigenvalues $\pm \sqrt{1-\vert z \vert^{2}}$, respectively.
Let $H:= \sum_{i=1}^{N}A^{(i)}\otimes \mathbb{I}^{N\setminus \lbrace i \rbrace}$ with \begin{eqnarray} A^{(i)} &=& \ket{\xi_{+}}\bra{\xi_{+}}-\ket{\xi_{-}}\bra{\xi_{-}} \nonumber \\ &=& {1\over \sqrt{1-\vert z \vert^{2}}}(\ket{\phi}\bra{\phi} - U\ket{\phi}\bra{\phi}U^{\dagger}) \end{eqnarray} for all $i$. $A^{(i)}$ clearly has unit operator norm, and so $H$ is a valid operator for optimization in the definition of $N^{rF}$ for spin systems. Because the pure state $\ket{\psi} \propto (\mathbb{I} + U)\ket{\phi}$ satisfies $\langle \psi \vert (\ket{\xi_{+}}\bra{\xi_{+}}-\ket{\xi_{+}}\bra{\xi_{+}}) \vert \psi \rangle = 0$, it follows that $\langle \Psi \vert H \vert \Psi \rangle = 0$. Hence the variance of $H$ in $\ket{\Psi}$ is then obtained by calculating the expectation value in $\ket{\Psi}$ of \begin{eqnarray}
H^{2}& =& \sum_{i\neq j} A^{(i)} \otimes A^{(j)}\otimes \mathbb{I}^{\otimes N \setminus \lbrace i,j \rbrace } \nonumber \\ & +& \sum_{i=1}^{N}A^{(i)2} \otimes \mathbb{I}^{\otimes N\setminus \lbrace i \rbrace }
\end{eqnarray} which is a simple computation. This result gives the formula in Eq.(\ref{eqn:variance}).

\section{\label{sec:povms}Equivalences among optimal distinguishing POVMs}

The results of Section \ref{sec:photons} allow us to demonstrate an equivalence between certain POVMs which are optimal for distinguishing the branches of the example photonic superpositions that were discussed in the text. The optimal POVM  $\lbrace P_{\pm} = (1/2)(\ket{\psi_{+}} \pm \ket{\psi_{-}})(\bra{\psi_{+}} \pm \bra{\psi_{-}}) \rbrace$ providing the maximal probability of successfully distinguishing $\ket{\alpha}$ and $\ket{-\alpha}$ in the Hilbert space $\mathcal{K} = \text{span} \lbrace \ket{\psi_{+}}, \ket{\psi_{-}} \rbrace \cong \mathbb{C}^{2}$ gives the same measurement results as the following POVM $\lbrace \tilde{P}_{\pm} \rbrace$ on $\mathcal{K}$:
\begin{widetext}
\begin{equation} \lbrace \tilde{P}_{\pm} := P_{\mathcal{K}}\left( {a^{2}e^{-2i\text{Arg}(\alpha)}+a^{\dagger 2}e^{2i\text{Arg}(\alpha)} \over 4\vert \alpha \vert^{2}} \pm \sqrt{1+e^{-4\vert \alpha \vert^{2}}}{a e^{-i\text{Arg}(\alpha)}+a^{\dagger}e^{i\text{Arg}(\alpha)} \over 4\vert \alpha \vert } \right) P_{\mathcal{K}} \rbrace
\end{equation}
\end{widetext}

Note that this compression of observables to the subspace $\mathcal{K}$ does not define a POVM on $\mathcal{H} = \ell^{2}(\mathbb{C})$.

Likewise, the optimal POVM $\lbrace P_{+} = \ket{n}\bra{n} , P_{-} = \ket{0}\bra{0} \rbrace $ for distinguishing $\ket{0}$ from $\ket{n}$ gives the same measurement results as
the following POVM on $\mathcal{K}$: \begin{equation}
\big\lbrace \tilde{P}_{+}:= P_{\mathcal{K}}{a^{\dagger}a\over n} P_{\mathcal{K}} \,  ,  \, \tilde{P}_{-}:= P_{\mathcal{K}}- P_{\mathcal{K}} \left( {a^{\dagger}a\over n} \right) P_{\mathcal{K}} \big\rbrace
\end{equation} 

Finally, the same weak equivalence can be formulated for the POVM which is used for distinguishing the branches of $\ket{\text{HCS}_{N}(\alpha)}$. The optimal POVM for distinguishing $\ket{\psi_{\pm}}$ is clearly $\lbrace \ket{\psi_{\pm}}\bra{\psi_{\pm}} \rbrace$. Using the fact that in the subspace $\mathcal{K}$, we have $\ket{\psi_{\pm}}\bra{\psi_{\pm}} = {\mathbb{I}\pm\sigma_{z} \over 2}$, it is clear that the following POVM 
on the subspace $\mathcal{K}$ produces the same measurement results as $\lbrace \ket{\psi_{\pm}}\bra{\psi_{\pm}} \rbrace$: \begin{equation}
\big\lbrace \tilde{P}_{\pm} := {P_{\mathcal{K}} \over 2} \pm {P_{\mathcal{K}} e^{i\pi {a^{\dagger}}a} a^{2} + a^{\dagger 2}e^{-i\pi {a^{\dagger}}a}P_{\mathcal{K}}  \over 2\text{Re}(\alpha^{2})} \big\rbrace .
\end{equation}

\begin{acknowledgments}
We thank the Kavli Institute for Theoretical Physics for hospitality during the program ``Control of Complex Quantum Systems'' in 2013, where this work was initiated, and for supporting this research in part by the National Science Foundation Grant No. PHY11-25915.
This work  
was also supported by NSF Grant No. CHE-1213141. 
\end{acknowledgments}

\bibliography{squeezed_references.bib}

\end{document}